\documentclass[12pt,preprint]{aastex}
\newcommand{\object}{CXO CDFS J033225.3-274219}
\newcommand{\objecta}{CXO CDFS J033236.8-274407}
\newcommand{\objectb}{CXO CDFS J033213.3-274241}
\newcommand{\chandra}{$Chandra$ }

\newcommand{\pksiron}{PKS~2149$-$306 }
\newcommand{\etal}{{\it et al.} }

\shortauthors{Wang et al.}
\begin{document}

\title{A PUZZLING X-RAY SOURCE FOUND IN THE \chandra DEEP FIELD SOUTH}
\author{J. Wang\altaffilmark{1},
T. Yaqoob\altaffilmark{1,2},
G. Szokoly\altaffilmark{3},
R. Gilli\altaffilmark{1,4},
L. Kewley\altaffilmark{5},
V. Mainieri\altaffilmark{6},
M. Nonino\altaffilmark{7},
P. Rosati\altaffilmark{6},
P. Tozzi\altaffilmark{7},
W. Zheng\altaffilmark{1},
A. Zirm\altaffilmark{1},
and C. Norman\altaffilmark{1,8}
}
\altaffiltext{1}{Dept. of Physics and Astronomy, The Johns Hopkins University,
Baltimore, MD 21218}
\altaffiltext{2}{Laboratory for High Energy Astrophysics, NASA/GSFC, code 662,
Greenbelt, MD 20771}
\altaffiltext{3}{Max-Planck-Institut f\"{u}r extraterrestrische Physik, Postfach 1312, D-85741 Garching, Germany}
\altaffiltext{4}{Istituto Nazionale di Astrofisica (INAF) - Osservatorio Astrofisico di Arcetri, Largo E. Fermi 5, 50125 Firenze, Italy}
\altaffiltext{5}{Harvard-Smithsonian Center for Astrophysics, 60 Garden Street, Cambridge, MA 02138}
\altaffiltext{6}{European Southern Observatory, Karl-Schwarzschild-Strasse 2, Garchin
g, D-85748, Germany}
\altaffiltext{7}{Istituto Nazionale di Astrofisica (INAF) - Osservatorio Astronomico, Via G. Tiepolo 11, 34131 Trieste, Italy}
\altaffiltext{8}{Space Telescope Science Institute, 3700 San Martin Drive,
Baltimore, MD 21218}
\vspace{0.1cm}

\begin{abstract}

In this letter we report the detection of an extremely strong X-ray 
emission line in the 940ks \chandra ACIS-I spectrum of \object.
The source was identified as a Type~1 AGN at redshift of $z$ = 1.617,
with 2.0 -- 10.0 keV rest frame X-ray luminosity of 
$\sim$ 10$^{44}$ ergs s$^{-1}$. The emission line was detected at 
6.2$^{+0.2}_{-0.1}$ keV, with an equivalent width (EW) of 4.4$^{+3.2}_{-1.4}$
keV, both quantities referring to the {\it observed frame}. In the rest 
frame, the line is at 16.2$^{+0.4}_{-0.3}$ keV with an
EW of 11.5$^{+8.3}_{-3.7}$ keV. An X-ray emission line at similar 
energy ($\sim$ 17 keV, 
rest frame) in \pksiron was discovered before using ASCA data. 
We reject the possibility that the line is due
to a statistical or instrumental artifact.
The line is most likely due to blueshifted Fe-K
emission from an relativistic outflow, probably an inner X-ray jet, 
with velocities of the order of $\sim 0.6-0.7c$.
Other possible explanations are also discussed.
\end{abstract}

\section{Introduction}

Along with the 1 Ms $Chandra$ exposure of 
\chandra Deep Field North (CDFN, Brandt et al. 2001, 2 Ms exposure
available now, see Alexander et al. 2002), 1 Ms exposure of 
\chandra Deep Field South (CDFS, Rosati et al. 2002) constitute the 
deepest X-ray exposure of a part of the sky ever taken.
With these two deep surveys,
the original X-ray glow discovered 
by Giacconi et al. in 1962
is now almost completely resolved into individual sources.
In CDFS, 346 such X-ray sources have been detected
in $\sim$ 0.1 deg$^2$ (Giacconi et al. 2002), most of which are extragalatic
sources harboring supermassive black holes. 
Great interest is now focusing on studying the properties of these X-ray
sources and understanding their physical nature.
In this letter, we report the discovery of an extremely strong X-ray 
emission-line feature in one of the
CDFS sources (a type~1 active galactic nucleus, or AGN, at $z=1.617$).
The line is at 16.2$^{+0.4}_{-0.3}$ keV in the rest frame, with a huge 
equivalent width of 11.5$^{+8.3}_{-3.7}$ keV.  
Both of these parameters are highly unusual. 

\section{The Data}
The 1~Ms exposure of CDFS
was composed of eleven individual \chandra ACIS observations from
Oct. 1999 to Dec. 2000. 
The detailed X-ray data reduction
and the final catalog of the X-ray detected sources were given in 
Giacconi et al. (2002).
\object, 6$\arcmin$ from the center of the field, was covered by all 
eleven exposures.
Its X-ray radial intensity profile
is consistent with that of a point-source 
(Giacconi et al. 2002), and no other X-ray source was detected within
38$\arcsec$.

The source was covered by the deep $R$ 
band [$R$(Vega)$<$26] primary optical imaging survey of CDFS,
which was obtained using the FORS1 camera on the ANTU (UT-1 at VLT) telescope.
The $R$ band image cut-out of the source overplotted with X-ray flux
contours is given in Figure~1. 
From the cut-out we can clearly see that the optical counterpart of 
the X-ray source is located at the center of 
the X-ray flux contours. The nearest other optical source is 5.6$\arcsec$ 
away, far outside of the 3$\sigma$ positional error circle (which has
a radius of 1.55$\arcsec$, see Giacconi et al. 2002 for details).
The CDFS was observed with the VLA at 1345 MHz and 4860 MHz in Oct,
1999 and Feb, 1999 respectively (Kellermann et al. in preparation).
The measured flux densities at the position of CXO CDFS
J033225.3-274219 at the two frequencies are 71 $\pm$ 13 and 43 $\pm$ 12
microJy corresponding to a spectral index of -0.41.
The source is radio loud with radio to optical ratio R = 160.
 
Low-resolution optical spectroscopy of \object\ was obtained with the
robotic multi-slit mode of FORS-1 on the ESO Very Large Telescope 8.2m
unit facility (ESO/VLT-ANTU) in November 23, 2000. A 1.2\arcsec\ slit
was placed on the object (seeing was 0.6\arcsec\ during most of the
observation). We used the lowest resolution, I150 grism and {\em no}
order-separation filter. The total integration time was 2.5 hours under
photometric conditions. The data were reduced following `standard industry
practices' and is shown in Figure~1. The object was classified
with relatively high confidence as a $z=1.617\pm0.01$ type-I AGN, 
based on two certain and one marginal broad emission line (C-III, Mg-II, C-IV,
with FWHM $>$ 1800 km/s).
The source has
an optical-to-X-ray ratio $\alpha_{ox}$ of 1.2, which is right of the
average value of soft X-ray selected AGNs (Puchnarewicz, et al. 1996).

In Figure~2, we plotted the X-ray count rates of the source extracted from
eleven $Chandra$ exposures. Since each exposure has similar effective area
(good to within $\pm$ 3\%) at the source position, the count rates
are representative of relative X-ray flux levels.
We can see that the continuum count
rates of all but one of the exposures are consistent with the average value
(within 3$\sigma$ errorbars).
We also looked into the individual exposures, but due to the limited counts,
no evidence of variance on a smaller time scale could be
established.

\section {X-ray spectral fitting}

The \chandra ACIS X-ray spectrum of the source was extracted from 
a circle with radius of 6.2$\arcsec$,
which is the 95\% encircled-energy radius of the
ACIS point-spread function at the source position,
and the background was extracted from an annulus with outer radius of 
18$\arcsec$ and inner radius of 8.2$\arcsec$. 
The X-ray telescope response (ARF) and CCD ACIS-I instrument response
was generated for each single \chandra observation,
and the final time-weighted files were used for spectral analysis.
The summed spectrum (source plus background) and the expected
background were shown in Figure~3.
After subtracting the background, we obtained around 440 X-ray counts in
the soft band (0.5 -- 2.0 keV), and 130 in the hard band (2.0 -- 9.0 keV).
Since we don't have enough counts, especially in the hard band,
we use C statistics (Cash 1979, Nousek \& Shue 1989) but compare with
results using $\chi^2$ for reference,
and XSPEC v11.2, which enables
the utilization of C statistics in the case when a background spectrum is 
also in use \footnote{see http://heasarc.gsfc.nasa.gov/docs/xanadu/xspec/manual/node57.html}, for the spectral-fitting analysis.
All the spectral fitting was done in the energy band  0.5 -- 9.0 keV.
All statistical errors
quoted in this paper are for 90\% confidence, one interesting parameter.

The unbinned spectrum was first fitted with a simple power-law plus a neutral 
absorber in the quasar frame. A Galactic 
neutral hydrogen absorption column of $8 \times 10^{19} \ \rm  cm^{-2}$
(Dickey \& Lockman 1990) was also included.
The X-ray spectrum is fairly steep ($\Gamma$ = 2.1$^{+0.2}_{-0.3}$) 
with moderate absorption
(N$_H$ = 1.3$^{+0.7}_{-0.8}$ $\times$ 10$^{22}$ cm$^{-2}$).
The absorption-corrected 2.0 -- 10.0 keV luminosity in the rest frame
is 0.8 $\times$ 10$^{44}$ ergs.s$^{-1}$ ($H_0$ = 60.0 km s$^{-1}$Mpc$^{-1}$, 
$\Omega_m$ = 0.3, $\Omega_\Lambda$ = 0.7).
The ratio of data to the model for the fit is given in Figure~4 (upper panel).
From both figure~3 and 4, we can see a huge bump at $\sim$ 6.2 keV (observed
frame), indicating a possible strong emission line. 
A single Gaussian was then added to our continuum model. 
With three more free parameters (the line centroid energy $E_c$, the line 
width $\sigma$, and 
the line intensity $I$), the fit was significantly improved 
($\Delta$ C = - 22.6), indicating a confidence level of 99.99\%.
The results are given in Table~1. 
In the quasar frame we obtain $E_c$ = 16.2$^{+0.4}_{-0.3}$ keV, $\sigma$ =
0.5$^{+0.6}_{-0.2}$ keV, and the  
equivalent width EW = 11.5$^{+8.3}_{-3.7}$ keV.
The large rest-frame energy and
large EW are both highly unusual and present a challenge
for the interpretation of the origin of the line emission.
We attempted to perform time-resolved spectroscopy
on the emission line but found that 
the data are not of sufficiently high quality
to investigate the line variability.

We also tried C and  $\chi^2$ statistics on the rebinned spectrum (at least 10
counts per bin), and present the results in table~1 for comparison.
We can see that three methods gave almost the same best-fit values for
the parameters, but different confidence levels.
Using the rebinned spectrum, C statistics gives higher significance
level of the line and slightly narrower errors than $\chi^2$ statistics does,
which is consistent with the prediction of Cash (1979).
Since rebinning the spectrum would smooth the line feature,
it's also reasonable that the line is more significant in the
unbinned spectrum than in the rebinned one. 
\section{Discussion}
\label{lineorigin}

It is clear that the emission line is not an artifact
of calibration uncertainty in the ACIS instrumental 
response function.
No such artifact (i.e. an emission-like feature at 
$\sim 6.2$ keV) has been reported, as far as we are aware.
In addition, we examined the ratios of data to a simple power-law
plus neutral absorber model for \objecta, and \objectb, two nearby sources 
with similar X-ray counts. These are also shown in Figure~4 for comparison, 
and it is clear that the feature is not observed in the other two sources.
Also, since the total exposure 
was composed of eleven individual observations
with different roll angles, photons from each source fell into
different positions on the instrument during different observations.
It is unlikely that an unknown instrumental artifact
(if there is any),  which occurs
only at certain positions on the
detector, would affect photons 
from all eleven exposures. Figure~2 shows the 5.7 -- 6.7 keV count rates
of \object\ from the eleven exposures, which are consistent with 
the average value within 3$\sigma$ errorbars. 
We note that the 5.7 -- 6.7 keV count rates from exposure No.1 and 5 are about
3$\sim$4 times of the average value. However, they only contribute 6\% of the
total exposure time and 19\% of the 5.7 -- 6.7 keV counts, and fittings with
the two exposures excluded yield similar results for the line
feature ($E_c$ = 6.1$^{+0.5}_{-0.2}$ 
keV, EW = 3.1$^{+3.8}_{-1.9}$ keV, both in the observed frame). 
This rules out the possibility that the line is due to an unknown artifact
which only affects one or several of the exposures. 
For the same reason, the line cannot either be due to a transient event
not associated with the source.
We can also eliminate inadequate background subtraction as the
origin of the emission line.
We tried using background spectra obtained from different
regions of the detector and obtained similar results.
An even more compelling reason 
is that the spectrum {\it without} background subtraction
still exhibits the emission line, but the background spectrum itself does not. 
And again the fact that we do not see the emission line in the
other two sources shown in Figure~4 
indicates that it is not due to the instrument
background or X-ray background.

Yaqoob et al. (1999) reported the discovery of a highly Doppler
blueshifted Fe-K emission line in QSO PKS 2149-306 from an ASCA
observation. In the quasar frame, the line has a similar center
energy (17.0$\pm{0.5}$ keV) with that of our source, but lower
equivalent width (300$\pm{200}$ eV). Furthermore, if we consider
both the lines as blueshifted Fe-K fluorescent lines,
the blueshift factors are also similar with that of 
another emission line detected in PKS 0637-75 at 1.6 keV 
(with EW = 60 eV, rest frame) from ASCA data, which was
interpreted as highly blueshifted O$_{VII}$ (Yaqoob et al. 1998).
The good agreements of the blueshift factors
from different sources and different instruments
make them more convincing and interesting.

So, could the line we detected also be explained as a highly 
Doppler blueshifted Fe-K emission line?
The Doppler factor of 2.3--2.5 then implies bulk velocities
of $\sim 0.62-0.72c$ (head-on) must be responsible for the blueshift
since the rest-frame energy of the line must be in the range
6.4--6.97 keV (neutral to H-like Fe).
Inflow (observed from the
far side) can be ruled out quite
simply because the only way to produce Keplerian velocities
high enough is to approach a compact object to within
a few gravitational radii, but by that time the gravitational
redshift is so large that it prevents the net
blueshift attaining a high enough value. Exact calculations
show this to be the case even for a maximally rotating Kerr
hole in which the rotation can help increase the blueshift.
Therefore, outflow is required. 
Yaqoob et al. (1999) pointed out that Fe-K line photons 
originating in an accretion disk and Compton scattering off a leptonic
jet aligned along the disk axis can account the emission line
in PKS 2149-306. However, this model can not produce the large
equivalent width we observed. Thus Fe-K line originating from
the outflow itself is required, and, to produce the line, the 
outflow can not be fully ionized.  Even then, if the emission
line is produced by fluorescence, the continuum photons
cannot illuminate the outflowing material from behind
since the Doppler redshift of the continuum would make it even harder
than it already is to
attain the large equivalent width. For the same reason,
there must be preferential blueshifting of the line photons
over the continuum photons. The continuum photons must
illuminate the outflow from the sides. 
So the outflow must locate very close to the X-ray continuum
emission region, and have comparable size to ensure the outflow
has enough open angle to the X-ray continuum emission to produce
the large equivalent width.
Under these conditions
it is possible to boost the rest-frame equivalent width
by the Doppler factor raised to the power of $3+\Gamma$
($\Gamma \sim 2.2$ is the X-ray photon index), or $\sim 2.5^{5.2}
\sim 117$.
Thus, an emission line which would ordinarily have an
equivalent width of 0.1 keV could be boosted to have
a rest-frame equivalent width of $\sim 12$ keV. 
Of course, the geometry is highly constrained.
We note that Fe-K lines have now been found in several
gamma-ray burst afterglows (e.g. Piro \etal 1999, 2000; Yoshida \etal 1999;
Antonelli \etal 2000) and some models
to account for these lines invoke super-solar Fe
abundances of as much as a factor $\sim 60$. If there is
any association at all with past gamma-ray bursts, or
if such super-solar abundances occur under other
circumstances, then the large equivalent width of the
emission line 
would obviously also be much
easier to explain (although an outflow is still required).

Similar blueshifted and strong iron emission line has not
been detected before in other AGNs. However, it's interesting to
note that Migliari et al. (2002) reported the discovery of 
blueshifted and very strong iron emission lines (EW = 13 keV, 
Migliari, private communication) from extended X-ray emission
in x-ray binary system SS 433. The X-ray jets are required to 
reheat itself to produce the X-ray continuum and line emission.
We argue that if the possible outflow of \object\ discussed above was also
heated by itself or other unknown mechanisms and ionized, it could 
produce iron line itself without being illuminated by the X-ray continuum.
Then the special geometry requirement can be released. 
It's interesting to note that VLA observation of the source resolved
an extended radio emission 6\arcsec\ south to the core
(Kellermann, private communication), presenting a possible radio jet.
Based on above discussions,
we conclude the most likely outflow is an inner X-ray jet. It's highly
valuable to study the source in more details in near future, including obtaining optical
spectroscopy with high quality to check if there are blueshifted optical
emission lines, obtaining X-ray image with better spatial resolution to
check if there is any extended X-ray emission, etc.

An alternative explanation is that some fraction of the X-ray 
continuum of \object\ is heavily absorbed by a cold outflow, and
the observed big bump is due to the Fe absorption edge.
The spectral fitting results indicate we need a cold outflow with  
bulk velocities of 0.74c and column density of
N$_H$ = 1.6 $\times$ 10$^{24}$ cm$^{-2}$, covering 98\% 
of the X-ray continuum ($\Delta$ C = - 17.9).
Such a cold outflow is unusual either.

Is it possible that there happens to be a low-redshift ($z \sim 0.034$)
type~2 AGN so close to \object\ that we cannot detect it individually? 
Fe K lines at 6.4 keV with large equivalent widths ($>$ 1 keV) are not
difficult to produce theoretically by Compton-thick type~2
AGNs (e.g. Ghisellini \etal 1994), and already confirmed by X-ray
observations (e.g. Levenson \etal 2002).
NGC 6240, a standard Compton-thick type~2 AGN, has an Fe K line intensity 
of $\sim$ 3.2
$\times$ 10$^{-5}$ photons cm$^{-2}$  s$^{-1}$ (Ikebe et al. 2000),
and a R band nuclei magnitude of 16.1 (Fried \& Schulz 1983).
Assuming a spectral energy distribution of NGC 6240, the iron line
flux we observed leads to a type~2 AGN with R = 22.5, which is even
brighter than \object\ (R=22.9). However, we didn't detect any evidence of 
its existence from either the deep $R$ image or the optical spectroscopy.
Even if the source was much more heavily obscured in optical than NGC 6240,
it is also puzzling that we would have missed its host galaxy (which has to 
have a low redshift, z $\sim$ 0.034) from our deep $R$ image.
Thus it is very unlikely that the emission line feature at 6.2 keV is
produced by another low redshift type~2 AGN. However, it
does not hurt to make a wild guess: is it possible that there is 
a class of source consisting of an obscured Compton-thick
active nucleus whose host galaxy is also obscured so that the entire object 
is optically invisible?
If this kind of source really exists, there might be more among the X-ray 
sources with no optical counterparts in the \chandra deep fields.

In principle there is a simple way to distinguish the blueshifted Fe
K line model from the outflowing absorption model or the type~1 plus 
type~2 model. 
That is, the latter two models predict that the X-ray emission above 
10 keV should be dominated by the outflow obscured component or the type~2 
AGN since the effect of the obscuring matter is much less than it is 
below 10 keV.
In fact using the latter two model we predict that the 10--20 keV flux 
should be $\sim 13$ times larger compared to the blueshifted Fe K line
model.
The planned hard X-ray imaging mission,
InFoc$\mu$s would be able to perform the simple test outlined here.

\acknowledgments
We would like to thank T. Heckman, J. Krolik, C. Reynolds, Y. Lu for discussions. We also thank the referee for a helpful report, especially on the use of C statistics which significantly improved the presentation of this letter.
\clearpage
\begin{deluxetable}{lccc}
\tablecaption{Spectral fits to \object}
\tablecolumns{4}
\tablewidth{0pt}
\tablehead
  {
  \colhead{Parameter} & \colhead{C} & \colhead{C} &\colhead{$\chi^2$}\\
  & \colhead{unbinned} & \colhead{rebinned} & \colhead{rebinned}
  }

\startdata

$\Gamma$  &  2.2$^{+0.2}_{-0.1}$   &  2.3$^{+0.3}_{-0.3}$ & 2.3$^{+0.3}_{-0.3}$ \\
$N_H$(10$^{22}$cm$^{-2}$) & 1.4$^{+0.4}_{-0.3}$ & 1.6$^{+1.0}_{-0.8}$ & 1.7$^{+1.2}_{-1.0}$ \\
$E_c$ (keV) &  6.2$^{+0.2}_{-0.1}$ & 6.2$^{+0.3}_{-0.2}$ & 6.2$^{+0.3}_{-0.2}$  \\
$\sigma$ (keV) & 0.2$^{+0.2}_{-0.1}$ & 0.2$^{+0.3}_{-0.2}$ & 0.2$^{+0.4}_{-0.2}$ \\
EW (keV) & 4.4$^{+3.2}_{-1.4}$ & 4.9$^{+3.5}_{-2.7}$ & 5.3$^{+3.5}_{-3.0}$ \\
$I$ & $1.3^{+1.0}_{-0.4}$ & $1.2^{+1.0}_{-0.6}$ & $1.3^{+0.8}_{-0.7}$ \\
C($\chi^2$)/dof & 548.8/575 & 48.8/49 & 46.7/49 \\
$\Delta$C ($\Delta \chi^2$)  & -22.6 & -16.1 & -11.4
\tablecomments{
Spectral fits with an absorbed power law plus Gaussian line model. 
The line intensity $I$ is in units of $10^{-7} \ \rm photons \ cm^{-2} \ s^{-1}$,
and the line parameters are given in the observed frame.
$\Delta$C ($\Delta \chi^2$) gives the improvements of the fit
by adding the Gaussian line to the continuum model.
}
\enddata
\end{deluxetable}

\newpage 
\begin{figure}
\plotone{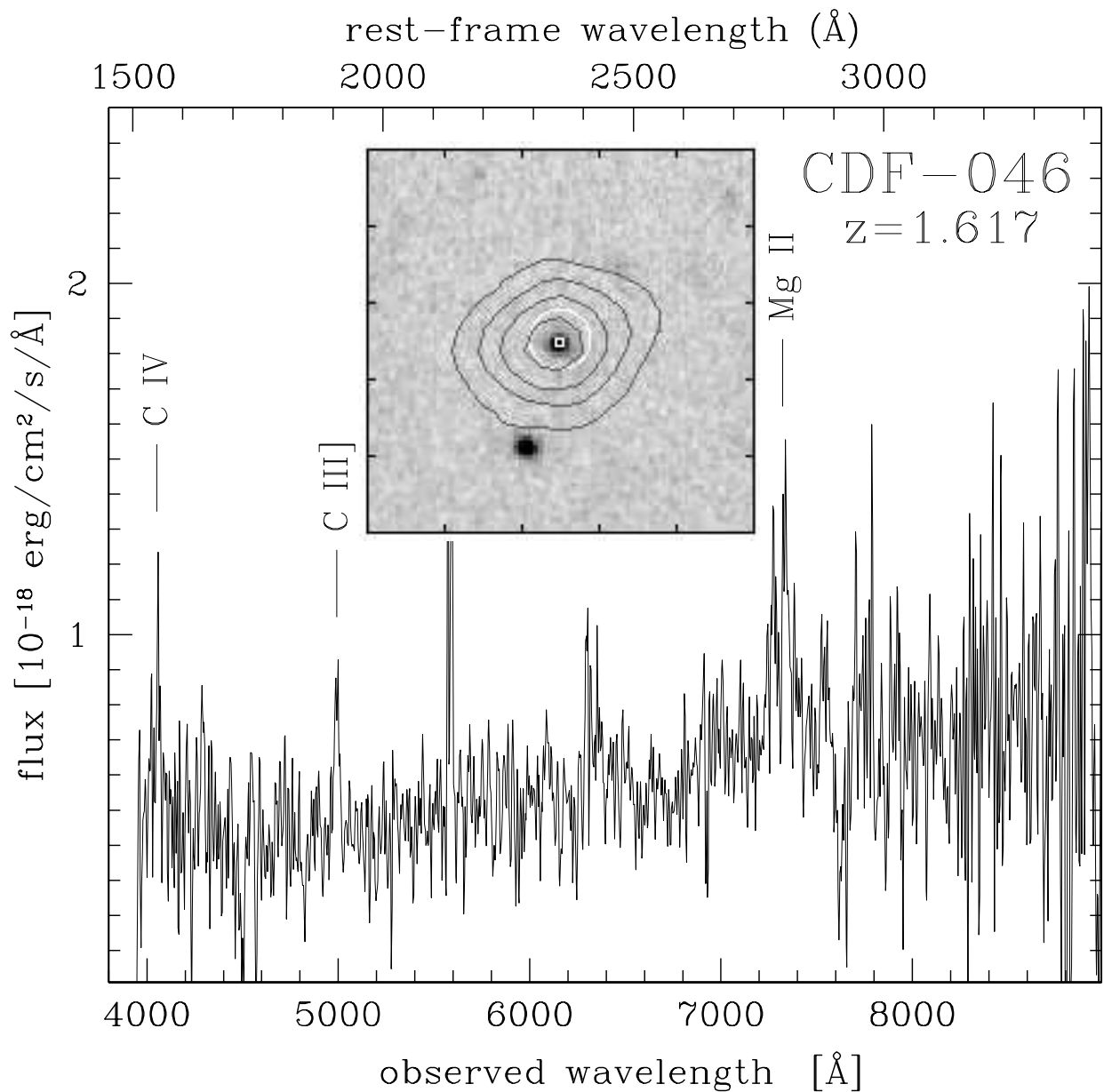}
\caption{
Low resolution optical spectroscopy of \object. 
R band cut-out is
also shown in the figure. The iso-intensity X-ray contours are at
2, 5, 10 and 20 sigma levels above the background. The size of the cutout
is 20$\arcsec$. The small circle in the cutout shows the 3$\sigma$
positional error of the X-ray source, and the optical counterpart was 
marked by a box.
}
\end{figure}

\begin{figure}
\plotone{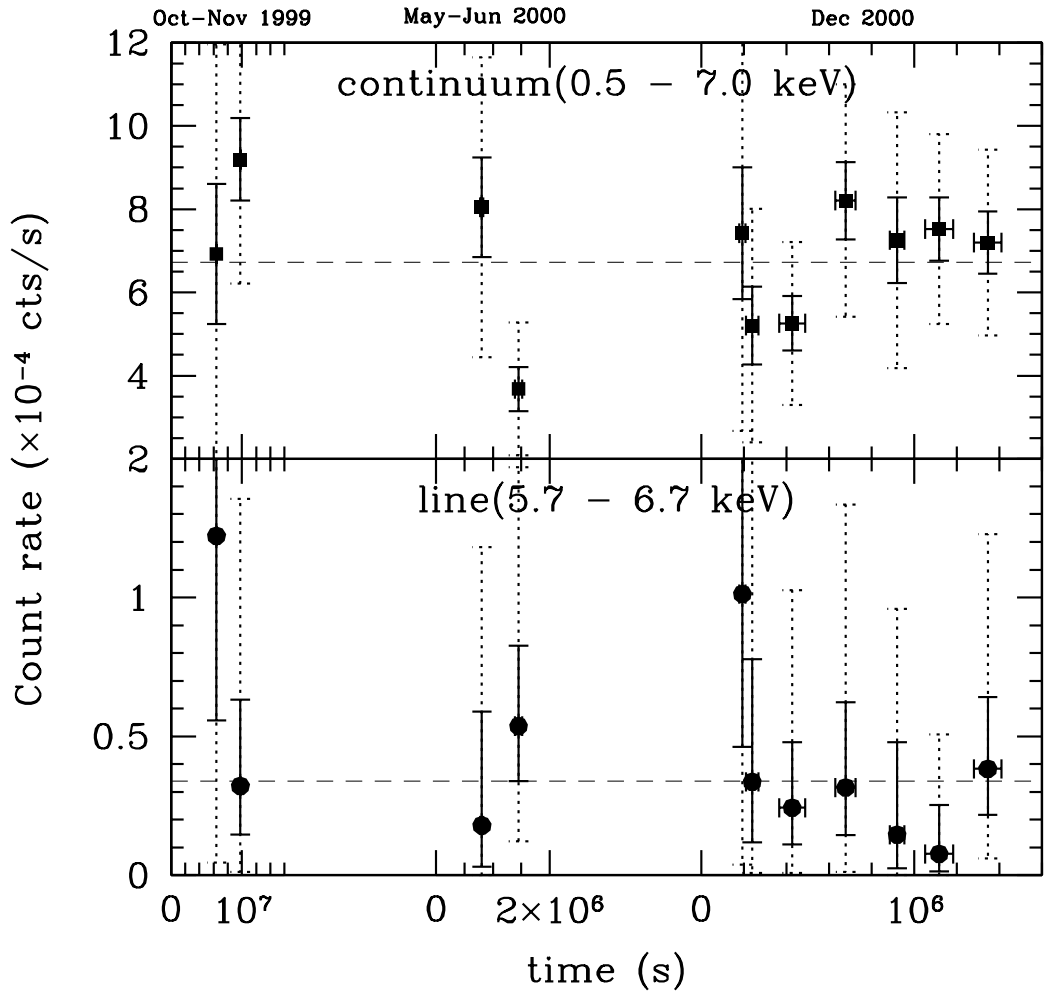}
\caption{
0.5 -- 7.0 keV and 5.7 -- 6.7 keV band count rates (source plus background) 
from 11 $Chandra$ exposures
in time sequence. Note that in order to show them in one figure, the X-axis
does not have a linear scale.
1$\sigma$ errorbars are shown in solid lines, and 3$\sigma$ errorbars shown in dotted lines.
Dashed lines are at the average value of the count rates.
}
\end{figure}

\begin{figure}
\plotone{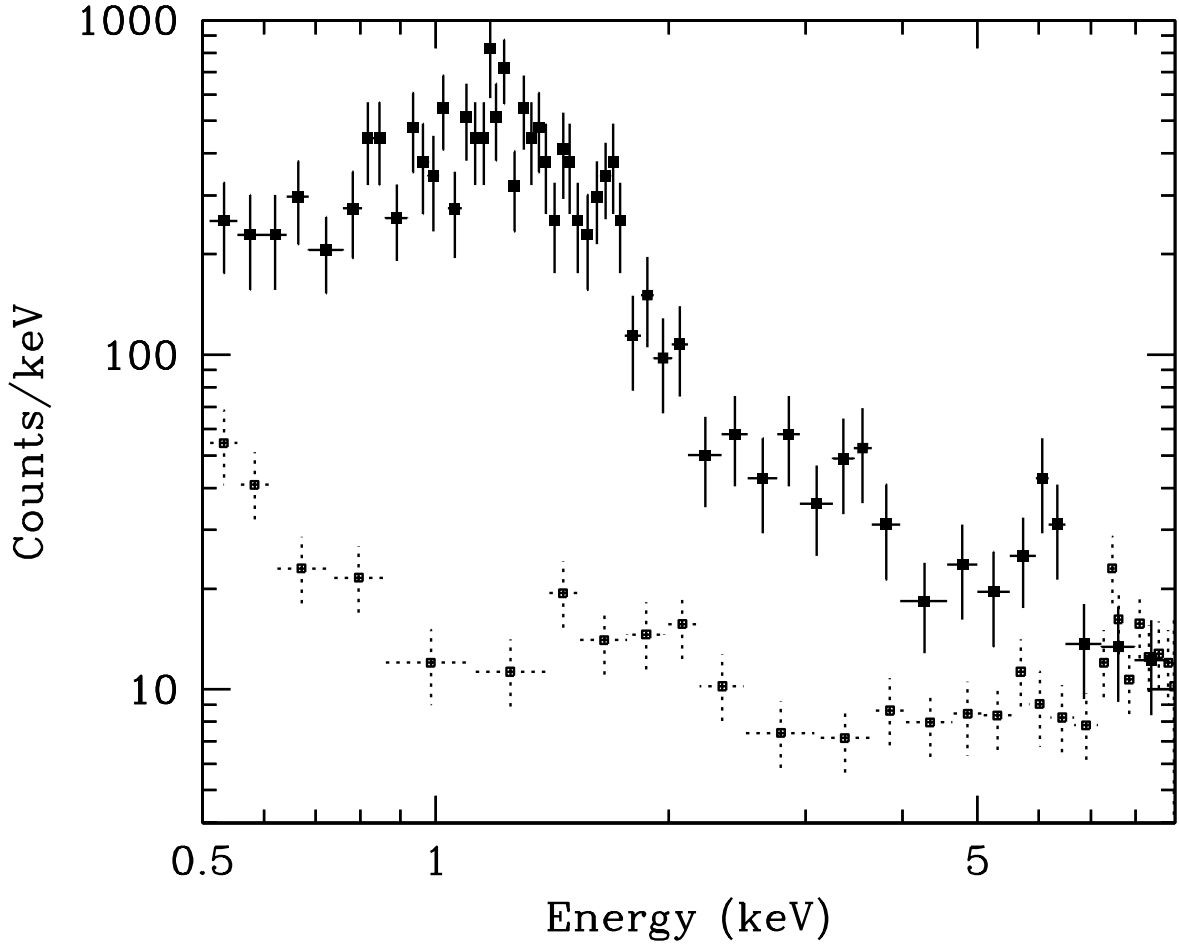}
\caption{
The summed (source plus background) X-ray spectrum of \object\
and the expected background (points with dotted error bars).
The spectra were rebinned only for display purpose.
}
\end{figure}

\begin{figure}
\plotone{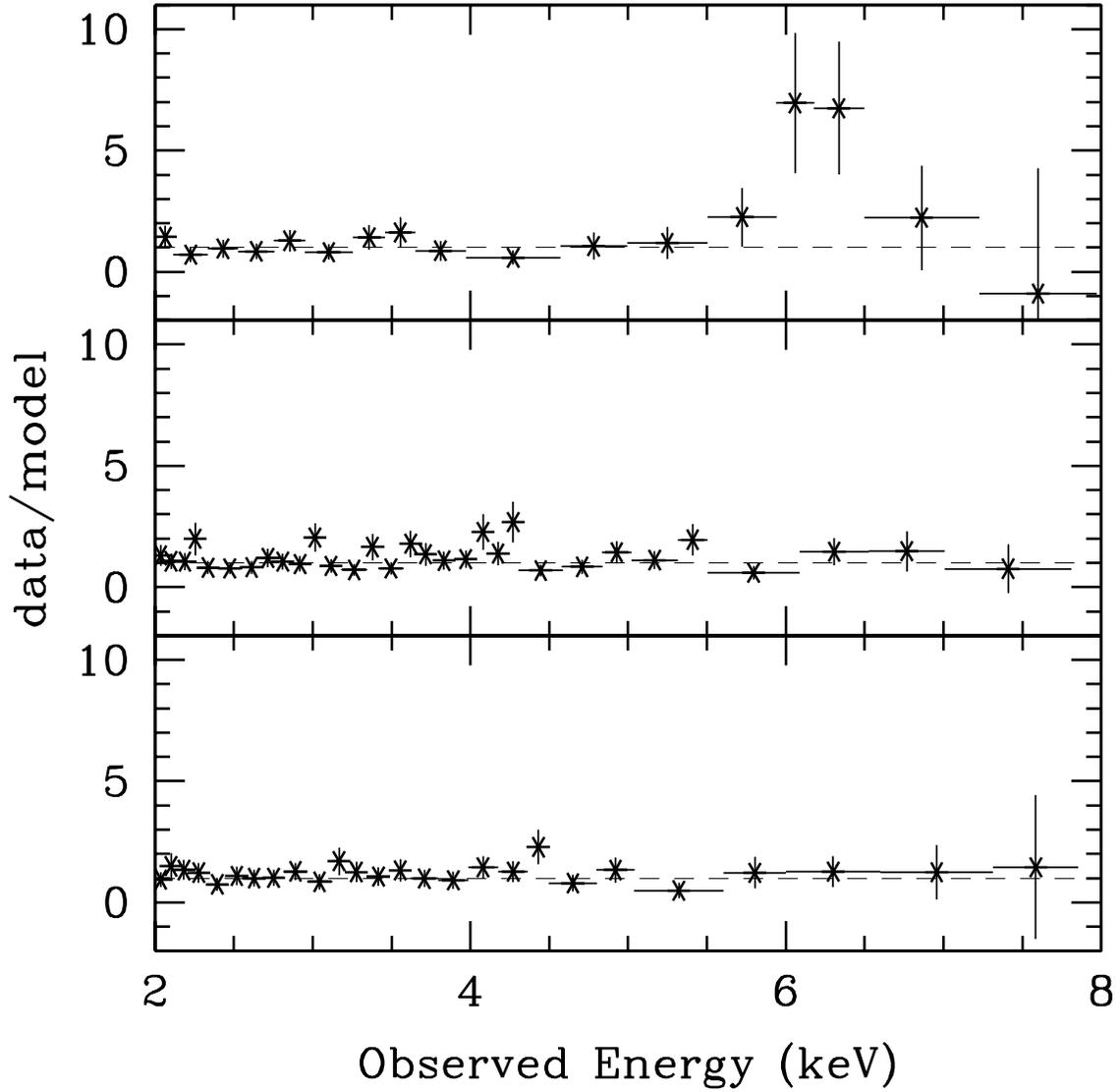}
\caption{
Ratios of \chandra ACIS spectral data to best-fitting power-law
plus neutral absorber models for upper panel: \object, and two nearby sources
with similar X-ray counts, middle panel: \objecta, and lower panel: \objectb.
The spectra were rebinned for display purpose.
}
\end{figure}

\end{document}